# Coherent cyclotron motion beyond Kohn's theorem

T. Maag[1], A. Bayer[1], S. Baierl[1], M. Hohenleutner[1], T. Korn[1], C. Schüller[1],

D. Schuh[1], D. Bougeard[1], C. Lange[1*], and R. Huber[1]

[1]*Department of Physics, University of Regensburg, 93040 Regensburg, Germany*

M. Mootz[2], J. E. Sipe[2,3], S. W. Koch[2], and M. Kira[2]

[2]*Department of Physics, University of Marburg, 35032 Marburg, Germany*

[3]*Department of Physics and Institute for Optical Sciences, University of Toronto, 60 St. George St.,*

*Toronto, Ontario, Canada*

**In solids, the high density of charged particles makes many-body interactions a pervasive principle governing optics and electronics[1-13]. However, Walter Kohn found in 1961 that the cyclotron resonance of Landau-quantized electrons is not influenced by the seemingly inescapable Coulomb interaction between electrons[2]. This surprising prediction has been confirmed by the strictly harmonic behavior of cyclotron oscillations which even persists in sophisticated quantum phenomena[14-17] such as ultrastrong light-matter coupling[18], superradiance[19], and coherent control[20]. Yet, the complete absence of nonlinearities excludes many intriguing possibilities, such as quantum-logic protocols[21]. Here, we use intense terahertz pulses to drive the cyclotron response of a two-dimensional electron gas beyond the protective limits of Kohn's theorem where anharmonic Landau ladder climbing and distinct terahertz four- and six-wave mixing signatures occur. Our theory explains these salient features as manifestations of dynamic Coulomb effects between the electrons and the positively charged ion background. These observations provide a new context for Kohn's theorem, unveiling previously inaccessible internal degrees of freedom of a Landau-quantized electron system, and open up new realms of ultrafast coherent quantum control for electrons.**

Controlling superpositions of electronic quantum states has been a paradigm of fundamental physics and quantum-information technology[21]. Sophisticated protocols have been implemented in atomic gases[22] whereas solids represent a more challenging environment due to complicated many-body interactions. Quantum manipulation has been quite successful in atom-like single-particle systems, such as quantum dots[23] while controlling interacting many-body systems is still in its infancy. Following Lev Landau's suggestion to combine collective properties of an interacting particle system into quasiparticles[1], researchers effectively analyse crystal electrons and holes[4,5], excitons[6,7], dropletons[12], polarons[8], magnons[24] or Cooper pairs[13]. In particular, ultrashort pulses in the terahertz (1 THz = $10^{12}$ Hz) spectral range have become a powerful tool to probe[4,7,8,10,13,19] and control[9,20,24,25] quasiparticle excitations, such as intra-excitonic transitions[6] or superconductor Higgs bosons[13].

For most quasiparticle excitations, Coulomb scattering leads to coherence times in the range of a few to a few hundred femtoseconds, which is too short for most quantum-logic operations. However, the cyclotron resonance (CR) in a two-dimensional electron gas (2DEG) represents a unique exception[2,14-20]. In a magnetically biased 2DEG, Landau electrons emerge as elementary quasiparticles. According to Kohn's theorem the quantisation energy of the harmonic spectrum of Landau levels (LLs) defined by the cyclotron frequency $\nu_c$ is inert to Coulomb scattering[2]. Although the absence of Coulomb complications warrants a long-lived CR coherence, the accompanying perfectly harmonic behavior has an apparent downside: Multiphoton excitations drive perfect Landau electrons only to climb up the LLs, preventing Rabi flopping, which is the desired elementary process in many quantum-logic operations[21].

Here, we investigate the possibility to induce a distinctly anharmonic CR response with the goal to coherently control transitions among few singled-out LLs. Intense THz pulses coherently drive the state of a 2DEG up the Landau ladder by as much as six rungs within a single cycle of the carrier wave and induce anharmonicities which render LL transitions distinguishable. Furthermore, strong coherent nonlinearities, such as four- and six-wave mixing signals are observed. Whereas such violations of Kohn's theorem could trivially result from nonparabolic electron dispersion[26] or impurities[27], our experiment–theory comparison unambiguously reveals that the nonlinear response is dominated by the Coulomb interaction between electrons and the positively charged ionic background.

Our sample hosts two 30-nm wide n-doped GaAs quantum wells (QWs) (see Methods), each containing a 2DEG. To calibrate the system, phase-locked low-amplitude ($\mathcal{E}_0 = 90$ V/cm) THz pulses are transmitted through the structure. Polarization components aligned parallel ($\mathcal{E}_x$) and perpendicular ($\mathcal{E}_y$) to the incident field are separately detected by electro-optic sampling (EOS), as a function of delay time $t$ (Fig. 1a). Without magnetic bias, $\mathcal{E}_x(t)$ (Fig. 1c, black curve) closely follows the incident waveform (Supplementary Discussion 1). When a magnetic bias $B = 3.5$ T is applied perpendicularly to the QW plane, long-lived trailing oscillations emerge due to the CR with $\nu_c = 1.45$ THz (Fig. 1c, blue curve)[20]. Here the weak THz field induces a polarization between LLs $|n = 0\rangle$ (filling factor $f = 0.95$) and $|n = 1\rangle$ (Fig. 1b), reemitting a circularly polarized THz wave (Fig. 1c, inset) whose coherence time $\tau_c = 9$ ps is likely limited by superradiant decay[19].

We then drive the calibrated system with intense THz pulses generated by tilted-pulse-front optical rectification[25]. The amplitude of the reemitted field $\mathcal{E}_y(t)$ (Fig. 2a) increases with $\mathcal{E}_0$, but the temporal waveform remains similar up to $\mathcal{E}_0 = 1.4$ kV/cm. For $\mathcal{E}_0 \geq 2.9$ kV/cm, the trailing oscillations slow down and decay more rapidly until the inter-LL polarization essentially follows the driving field, for $\mathcal{E}_0 = 8.7$ kV/cm, as the coherence decays almost instantly. The spectrum of the reemitted field (Fig. 2b), which is initially centered at $\nu_c = 1.45$ THz, red-shifts and broadens with increasing $\mathcal{E}_0$ until its shape converges to the spectrum of the driving pulse. This pronounced field dependence is irreconcilable with the linear response of a harmonic oscillator, indicating that the cyclotron transition is driven into a strongly nonlinear regime by the intense THz transients. Figure 2c shows the dephasing time $\tau_c$ as a function of the THz amplitude $\mathcal{E}_0$. Starting at a low-field value of $\tau_c = 9$ ps, the decay time drops abruptly for $\mathcal{E}_0 > 3$ kV/cm, and ultimately approaches $\tau_c = 0.5$ ps.

Kohn's theorem implies that the purely repulsive electron–electron Coulomb interaction cannot produce observable nonlinearities in a 2DEG. However, in a real modulation-doped system, the 2DEG always resides on a positive background charge that holds the electrons together as a homogeneous gas by preventing the formation of a Wigner crystal[28] or the escape of electrons to the edges of the sample. Since the attractive electron–ion interaction is not limited by Kohn's theorem, it can indeed induce CR nonlinearities. To explain the experimental observations, we perform a fully quantum mechanical many-body calculation, including both the electron–electron and electron–ion contributions. Longitudinal optical (LO) phonon scattering is described via a dephasing time of 9 ps (0.8 ps) for LLs with energies below (above) $\hbar\omega_{LO} = 36$ meV (Methods and Supplementary Discussion 4).

Figure 2d compares the transmitted field $\mathcal{E}_y$ for $\mathcal{E}_0 = 0.7$ kV/cm and 5.7 kV/cm as obtained from our full theory (solid curves) with the results of a calculation where the Coulomb interaction is switched off (dashed curves) and a classical calculation with a nonparabolic band and constant $\tau_c$ (shaded area). The weak-intensity result, here for $\mathcal{E}_0 = 0.7$ kV/cm, is well reproduced by all theoretical models, verifying the applicability of Kohn's theorem in this case. However, for the stronger field $\mathcal{E}_0 = 5.7$ kV/cm, the simplified calculations fail qualitatively. Only the full quantum theory can reproduce the experimental decay of $\mathcal{E}_y$ and the redshift of $\nu_c$. As in the experiment, the indicated oscillation minimum of $\mathcal{E}_y$ (red vertical line) is delayed by 0.1 ps with respect to the position expected for constant $\nu_c = 1.45$ THz (blue vertical line), which manifests a violation of Kohn's theorem. Even though the simplified models neglect Coulomb interaction, they do yield a clear change of $\nu_c$ implying that the nonparabolic dispersion contributes to the softening of the CR. However, the two models predict different decay dynamics because the classical computation lacks the phase diffusion among LLs, present in the quantum calculations.

Our full quantum calculations also explain the threshold-like onset of dephasing above $\mathcal{E}_0 = 3$ kV/cm (Fig. 2c, red solid line) attributing this effect to an efficient population transfer. While low THz amplitudes (0.7 kV/cm) excite only a few percent of charge carriers into LL $|n = 1\rangle$ (Fig. 2e, blue bars), strong pulses drive coherent ladder climbing. At $\mathcal{E}_0 = 4.3$ kV/cm (violet bars), more than half of the electrons are excited out of the ground state $|n = 0\rangle$ and distributed up to $|n = 6\rangle$ where rapid dephasing and relaxation by LO phonon emission sets in. Remarkably, THz transients with $\mathcal{E}_0 = 8.7$ kV/cm (red bars) largely depopulate $|n = 0\rangle$ and prepare a compact superposition of eigenstates, peaking at $|n = 5\rangle$. With 25% of its weight located above the phonon energy, this wavepacket rapidly loses its coherence. Thus, the abrupt onset of dephasing above $\mathcal{E}_0 = 3$ kV/cm arises from the combination of a compact wavepacket distribution and a sharp threshold for phonon scattering. Furthermore, the experiment corroborates that below the phonon threshold, coherent state inversion works even for a massive many-body system and corresponding LLs should be well-suited for coherent quantum control.

To test these perspectives systematically, two phase-locked THz field transients polarised in x and y-direction and labelled A and B, respectively, are focused onto the sample for field-resolved two-dimensional (2D) THz spectroscopy. The transmitted total THz field is electro-optically recorded with amplitude and phase resolution as a function of the EOS time $t$ and the delay between the two incident pulses, $\tau$ (Fig. 3a). Subsequently, the response induced individually by pulses A and B is subtracted to isolate the nonlinear polarization $\mathcal{E}_{\text{NL}}(t,\tau)$[29,30] which vanishes for a strictly harmonic CR. In contrast, Figs. 3b-e show strong nonlinear signals $\mathcal{E}_{\text{NL}}(t,\tau)$ for all peak amplitudes of pulse A between $\mathcal{E}_A = 1.4$ kV/cm and 5.7 kV/cm. The amplitude of pulse B is kept constant at $\mathcal{E}_B = 90$ V/cm. Constant-phase lines of pulse B appear vertically in the 2D data maps whereas the phase fronts of pulse A occur under a 45° angle (see Supplementary Discussion 3).

For $\mathcal{E}_A = 1.4$ kV/cm (Fig. 3b), $\mathcal{E}_{\text{NL}}$ is strongly modulated along $\tau$ with a period $\Delta\tau_1 = 1/\nu_c = 0.69$ ps evidencing that the nonlinear interaction is coherently mediated by $\nu_c$. This feature persists even when the incident pulses do not overlap in time since the coherence is stored in the LL system. For larger $\mathcal{E}_A$ (Figs. 3c-e), an additional modulation with a period of $\Delta\tau_2 = 2/\nu_c = 0.34$ ps, corresponding to twice the cyclotron frequency, emerges and becomes dominant for $\mathcal{E}_A = 4.3$ kV/cm. Increasing the field to $\mathcal{E}_A = 5.7$ kV/cm (Fig. 3e) accelerates decoherence, leaving only weak modulations during temporal overlap of both pulses, around $t + \tau = 0$.

A 2D Fourier transformation allows us to disentangle different nonlinear optical processes contributing to $\mathcal{E}_{\text{NL}}(t,\tau)$. Figures 3h-k display the spectral amplitude as a function of the frequencies $\nu_t$ and $\nu_\tau$ associated with the EOS time and the relative delay between the pulses, respectively[29,30]. Several distinct maxima occur at integer multiples of $\nu_c$: The peak located at $(\nu_t, \nu_\tau) = (\nu_c, 0)$ (Fig. 3h, arrow 'k$_{p1}$') represents a pump–probe signal where pulse A (B) acts as a pump (probe) pulse, whereas the maximum at $(\nu_c, -\nu_c)$ (arrow 'k$_{p2}$') is a pump–probe signal for which pulse A and B switch their roles. In addition, strong four-wave-mixing (FWM, 'k$_{41}$', 'k$_{42}$') emerges. As $\mathcal{E}_A$ is increased to 2.9 kV/cm (Fig. 3i), the FWM signal at $k_{42}$ (red circle) surpasses the diagonal pump–probe signal $k_{p2}$ in amplitude. Even six-wave mixing (SWM, 'k$_{61}$', orange circle) becomes clearly discernible. For $\mathcal{E}_A = 4.3$ kV/cm (Fig. 3j),

FWM is the dominant contribution at non-zero $\nu_\tau$ and explains the $\Delta\tau_2$-periodic response in the time-domain (Fig. 3d). Finally, the strongly reduced coherence time for $\mathcal{E}_A = 5.7$ kV/cm (Fig. 3k) suppresses wave-mixing processes, and the incoherent pump–probe signal at $k_{p1}$ dominates.

To identify the origin of these nonlinearities, we apply our many-body theory to the 2D scenario using the experimental THz waveforms. The time-domain data of the full calculation for weak ($\mathcal{E}_A = 1.4$ kV/cm, Fig. 3f) and strong ($\mathcal{E}_A = 4.3$ kV/cm, Fig. 3g) pulses explain our experiment, reproducing the $\Delta\tau_1$- and $\Delta\tau_2$-periods in $\mathcal{E}_{NL}(t,\tau)$, as well as their relative weights and decay. Correspondingly, the frequency maps (Figs. 3l, m) exhibit multi-wave mixing features in quantitative agreement with the experiment.

A switch-off analysis allows us to gauge the relative weight of microscopic nonlinearities. In Fig. 4, we plot slices of $\mathcal{E}_{NL}$ at constant $t = 3.3$ ps (see Figs. 3f, g, vertical lines). For $\mathcal{E}_A = 1.4$ kV/cm (Fig. 4a), the $\Delta\tau_1$-period dominates the dynamics of $\mathcal{E}_{NL}(\tau)$. When the nonparabolicity is switched off (red line), $\mathcal{E}_{NL}$ is strongly suppressed. Eliminating Coulomb effects (dashed line), in contrast, yields almost unchanged $\mathcal{E}_{NL}$, revealing the band structure as the main contributor to the nonlinearities for low fields. For $\mathcal{E}_A = 4.3$ kV/cm (Fig. 4b), the nonlinearities are almost one order of magnitude stronger than for $\mathcal{E}_A = 1.4$ kV/cm. Now the parabolic band approximation (red line) produces virtually the same result as the full computation (black line), showing that the Coulomb interaction dominates $\mathcal{E}_{NL}$ for strong fields. Additional detailed insight into the unusual quantum state of the strongly driven 2DEG is gained by computing $\mathcal{E}_{NL}(\tau)$ without excitation-induced dephasing (EID, dashed line, see Supplementary Discussion 10); only the computation with EID explains the experiment.

In summary, lifting the protection of Kohn's theorem, non-perturbative excitations of a Landau system open the door to a rich spectrum of nonlinearities that can be fine-tuned by the THz driving field. We have shown how transient nonequilibrium populations and strong polarization fields drastically alter the effectiveness of the Coulombic interaction between the 2DEG electrons and the ionic background. In this scenario, strong many-body interactions cause non-perturbative four- and six-wave mixing THz nonlinearities which stand in stark contrast to the linear response of the Landau quantum harmonic oscillator expected by Kohn's theorem. Many-body dynamics in a quantum system can thus be exploited to address internal degrees of freedom inaccessible for linear optics, facilitating sophisticated quantum control even in massive many-body systems.

**Methods**

**Experiment.**

Our sample, grown by molecular beam epitaxy, hosts two 30-nm wide GaAs quantum wells separated by a 10-nm wide $Al_{0.24}Ga_{0.76}As$ barrier. In the 2DEGs a carrier density $\rho_e = 1.6 \times 10^{11}$ cm$^{-2}$ is realized by two remote δ-doping layers, symmetrically enclosing the QWs on both sides. $Al_{0.24}Ga_{0.76}As$ spacers of 72 nm thickness serve as barriers between the QWs and the doping layers (see Supplementary Information 2). A high electron mobility of $\mu = 4.6 \times 10^6$ cm$^2$/Vs results.

A Ti:sapphire laser amplifier (centre wavelength: 800 nm, pulse energy: 5.5 mJ, repetition rate: 3 kHz, pulse duration: 33 fs) is used to generate intense few-cycle THz fields by tilted pulse-front optical rectification in a cryogenically cooled $LiNbO_3$ crystal. In a second optical branch, a small portion of the laser energy drives optical rectification in a 180-μm thick (110)-cut GaP crystal. Both THz pulses are collinearly focused onto the sample which is mounted in a magneto-optical cryostat at a constant temperature of 4.3 K. The superconducting split-coil magnet provides homogeneous fields which are polarized perpendicularly to the quantum well plane and tunable between 0 and 5 T. By varying the relative temporal delay between the maxima of the two transients with mechanical delay stages, we perform field-resolved two-dimensional (2D) THz spectroscopy (c.f. Supplementary Information 3 for a detailed discussion). The transmitted THz pulses pass a rotatable wire-grid polarizer and are focused onto a 0.5-mm thick (110)-cut ZnTe crystal for polarization-resolved electro-optic detection covering the frequency window between 0.1 and 3.0 THz.

**Theory.**

Our full many-body theory includes the electron–electron and electron–ion interaction as well as nonparabolic energy dispersion, as elaborated in Supplementary Discussion 4. The many-body dynamics is solved with the semiconductor Bloch equations (SBEs)[3] for a 2DEG in a static magnetic field where we account for LO phonon scattering as well as EID, presented in Supplementary Discussion 10 together with the explicit form of the SBEs. To obtain converged macroscopic responses, the SBEs are solved for $2.25 \times 10^6$ THz-induced transition amplitudes between the Landau levels. We propagate the related integro-differential equations with a fourth-order Runge–Kutta method, and each time step involves more than $10^{10}$ calculations due to the Coulomb coupling, which makes the simulations extremely demanding. Nevertheless, this formidable challenge is executable via an efficient parallel-computing implementation. To study the limitations of Kohn's theorem, we also perform a classical calculation with a nonparabolic energy dispersion and a constant decay, see Supplementary Discussion 4 pp. for a more detailed discussion.

**Acknowledgements** The work in Regensburg was supported by the European Research Council through grant no. 305003 (QUANTUMsubCYCLE) and the Deutsche Forschungsgemeinschaft (LA 3307/1-1, HU 1598/2-1, and BO 3140/3-1). The work at the University of Marburg was supported by the Deutsche Forschungsgemeinschaft through SFB 1083 and grant KI 917/2-2 (M.K.), and the Alexander von Humboldt foundation (J.S.).

**Author contributions** T.M., A.B. M.M. and C.L. contributed equally to this work. C.L., T.M., M.K., S.W.K. and R.H. conceived the study. T.M., C.L., A.B., S.B., M.H., T.K., C.S., D.B. and R.H. carried out the experiment and analysed the data. A.B., D.S. and D.B. prepared the sample. M.M., J. S., S.W.K. and M.K. developed the quantum-mechanical model and carried out the computations. T.M., C.L., M.M., S.W.K., M.K. and R.H. wrote the manuscript. All authors discussed the results.

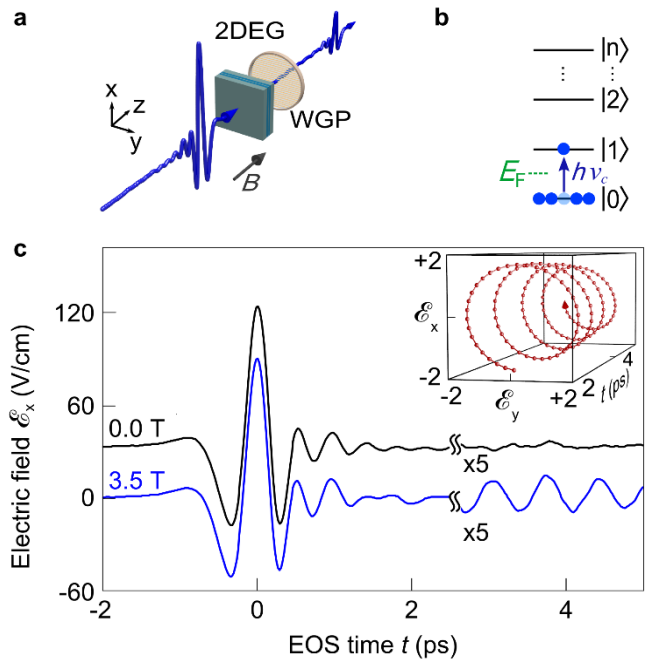

**Figure 1 | Linear THz magnetospectroscopy of a 2DEG Landau system. a**, Schematic of time-domain THz setup in Faraday geometry. Linearly polarized THz transients drive inter-Landau level transitions in two 30-nm wide, n-doped GaAs QWs. The transmitted waveforms are detected via electro-optic sampling (EOS) in a ZnTe crystal with polarization selectivity achieved through a wire-grid polarizer (WGP). **b**, Landau level scheme. At $B$ = 3.5T, all electrons condense to the first Landau state $|n = 0\rangle$ with a filling factor of $f = 0.95$. **c**, Transmitted THz pulses at $B$ = 3.5 T (blue curve) and $B$ = 0 T (black curve, offset for clarity). For $B$ = 3.5 T, trailing oscillations result from the decay of the polarization, which our low-field THz transient excites solely between LLs $|n = 0\rangle$ and $|n = 1\rangle$. Inset: Polarization-resolved plot of the reemission showing its circular nature.

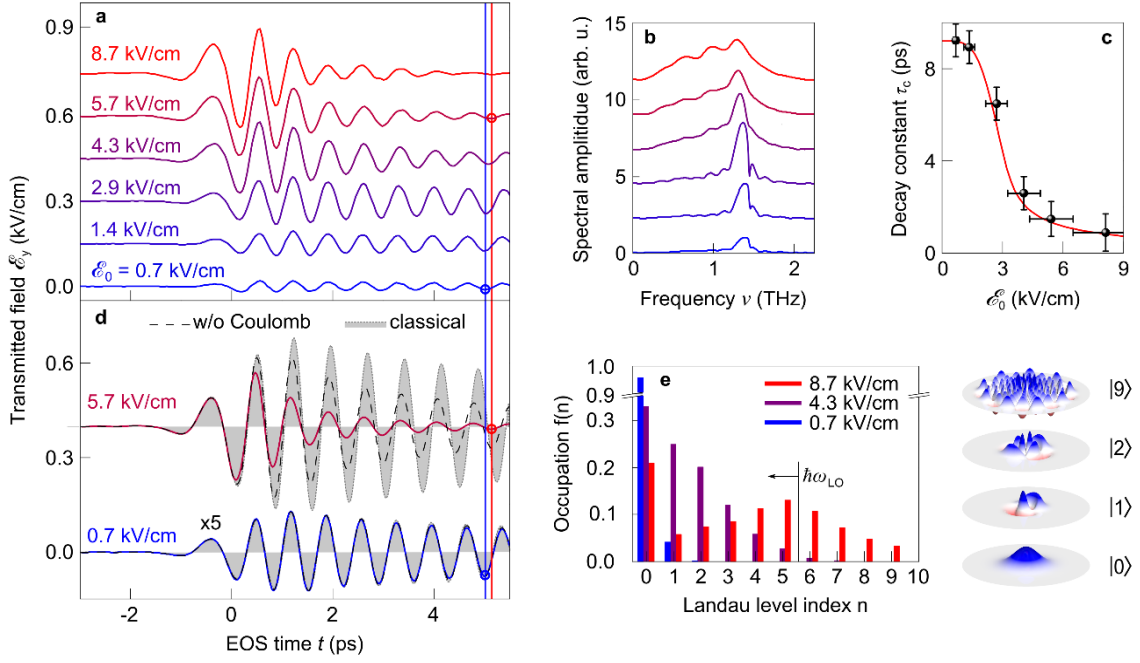

**Figure 2 | Dynamics of Landau system under nonperturbative single-pulse excitation. a**, Waveform of the transmitted field $\mathcal{E}_y$ for various amplitudes $\mathcal{E}_0$ of the incident THz transient (polarized in x-direction), vertically offset for clarity. Vertical lines indicate the temporal position of a local minimum of $\mathcal{E}_y$ for low-field excitation ($\mathcal{E}_0$ = 0.7 kV/cm, vertical blue line) and high-field excitation ($\mathcal{E}_0$ = 5.7 kV/cm, vertical red line), indicating a field-induced phase retardation. **b**, Fourier transform of $\mathcal{E}_y$ from **a** evidencing the broadening and red-shift of the cyclotron resonance from 1.45 to 1.3 THz. **c**, Coherence time $\tau_c$ extracted from fitting the waveforms $\mathcal{E}_y(t)$ in **a** with an exponentially decaying sinusoidal function (black dots), and calculation (red line). **d**, THz dynamics calculated with a full quantum theory for $\mathcal{E}_0$ = 0.7 kV/cm (blue curve) and $\mathcal{E}_0$ = 5.7 kV/cm (red curve). Dashed curves: quantum theory neglecting Coulomb interaction, shaded areas: classical theory including the band nonparabolicity of GaAs. **e**, Bar chart of calculated Landau level population as a function of LL index $n$ at a fixed delay time $t$ = 0.6 ps, for $\mathcal{E}_0$ = 0.7 kV/cm (blue bars), $\mathcal{E}_0$ = 4.3 kV/cm (violet bars) and $\mathcal{E}_0$ = 8.7 kV/cm (red bars). Phonon scattering sets in for energies above the LO phonon energy $\hbar\omega_{LO}$ (vertical black line and arrow). Right sketch: real part of Landau wavefunctions for $n$ = 0,1,2, and 9, with $l = n$.

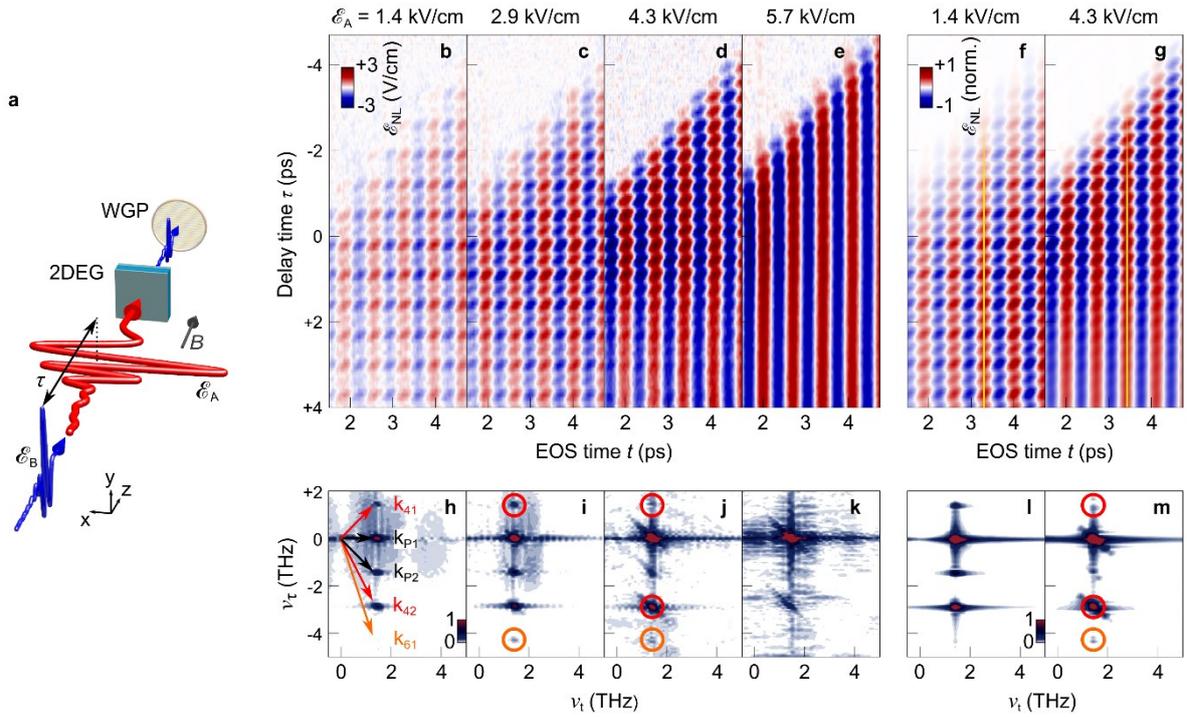

**Figure 3 | Two-dimensional, phase-resolved, collinear THz spectroscopy. a**, Schematic: Strong THz pulses A of amplitude $\mathcal{E}_A$ (polarized in x-direction) prepare a coherent inter-LL polarization in the 2DEG. The nonlinear THz response $\mathcal{E}_{NL}(t, \tau)$ is time-resolved using weak pulses B of amplitude $\mathcal{E}_B$ = 90 V/cm (polarized in y-direction), incident after a relative delay time $\tau$. **b**-**e**, $\mathcal{E}_{NL}(t, \tau)$ for $\mathcal{E}_A$ = 1.4 – 5.7 kV/cm. **h**-**k**, Two-dimensional Fourier transform of $\mathcal{E}_{NL}(t, \tau)$ from **b**-**e**. Labels and circles highlight spectral positions of coherent pump-probe ($k_{p1}$, $k_{p2}$, black), four- ($k_{41}$, $k_{42}$, red) and six-wave-mixing nonlinearities ($k_{61}$, orange), and arrows visualize wave vectors perpendicular to phase fronts of the respective processes. **f**,**g** and **l**,**m**, $\mathcal{E}_{NL}(t, \tau)$ and corresponding spectra calculated through a microscopic quantum theory for two representative amplitudes, $\mathcal{E}_A$ = 1.4 kV/cm, and $\mathcal{E}_A$ = 4.3 kV/cm. Vertical yellow lines indicate the sampling time of $t$ = 3.3 ps for which $\mathcal{E}_{NL}$ is analyzed in Fig. 4.

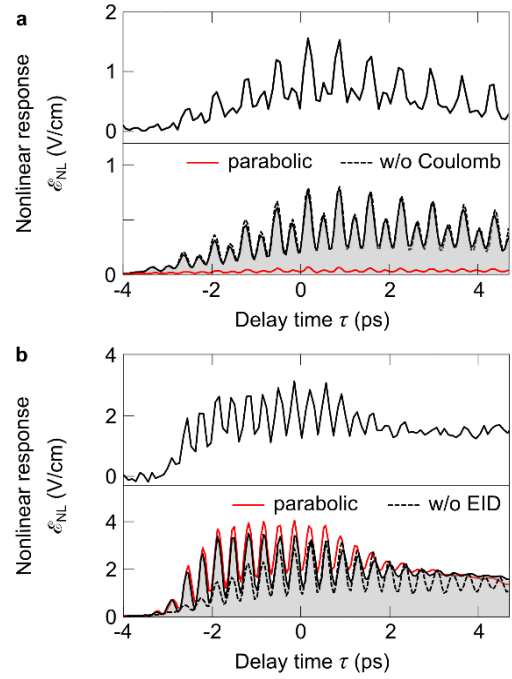

**Figure 4 | Time-domain switch-off analysis of nonlinear contributions. a**, Nonlinear response $\mathcal{E}_{NL}(t, \tau)$ for $\mathcal{E}_A$ = 1.4 kV/cm at a constant sampling time of $t$ = 3.3 ps as measured (top panel). Bottom panel: Calculation via a full quantum theory (black curve and shaded area). Dashed curve: calculation excluding Coulomb interaction. Red curve: calculation assuming a parabolic band. **b**, $\mathcal{E}_{NL}(t, \tau)$ for $\mathcal{E}_A$ = 4.3 kV/cm. Top panel: experiment. Bottom panel: full calculation (black curve and shaded area), calculation assuming a parabolic band (red curve), and calculation neglecting excitation-induced dephasing (dashed curve).